\documentclass[twocolumn,showpacs,preprintnumbers,amsmath,amssymb]{revtex4}

\usepackage{graphicx}
\usepackage{dcolumn}
\usepackage{bm}
\usepackage{color}

\begin{document}

\preprint{APS/123-QED}

\title{Maximum Likelihood and the Single Receptor}

\author{Robert G. Endres${}^{1,2,*}$ and Ned S. Wingreen${}^{3,\dagger}$}
 \affiliation{
   {${}^1$Division of Molecular Biosciences, Imperial College London, 
    London SW7 2AZ, United Kingdom}\\
   {${}^2$Centre for Integrated Systems Biology at Imperial College, Imperial College London, 
    London SW7 2AZ, United Kingdom}\\
   {${}^3$Department of Molecular Biology, Princeton University, Princeton, NJ 08544-1014}}

\date{\today}

\begin{abstract}
Biological cells are able to accurately sense chemicals with receptors
at their surfaces, allowing cells to move towards sources of attractant
and away from sources of repellent. The accuracy of sensing chemical concentration
is ultimately limited by the random arrival of particles at the receptors by 
diffusion. This fundamental physical limit is generally considered to be the 
Berg \& Purcell limit [H.C. Berg and E.M. Purcell, Biophys. J. {\bf 20}, 193 (1977)]. 
Here we derive a lower limit by applying maximum likelihood to the 
time series of receptor occupancy. The increased accuracy stems from solely 
considering the unoccupied time intervals - disregarding the occupied time 
intervals as these do not contain any information about the external particle
concentration, and only decrease the accuracy of the concentration estimate. 
Receptors which minimize the bound time intervals achieve the highest possible 
accuracy. We discuss how a cell could implement such an optimal sensing strategy 
by absorbing or degrading bound particles.

\end{abstract}


\maketitle

Single cells can sense external chemical concentrations with extremely high accuracy. For instance, 
the chemotactic bacterium {\it Escherichia coli} can detect $3.2$ nM of the attractant
aspartate \cite{manson03}, which corresponds to only 
about $3$ attractant particles in the volume of the cell. Single eukaryotic cells such as
{\it Dictyostelium discoideum} \cite{gradient_sensing} and  {\it Saccharomyces cerevisiae} 
\cite{segall93} (budding yeast) are well known to measure and respond to extremely 
shallow gradients of chemical signals \cite{levine}. 
These observations raise the question how close do cells operate to the fundamental physical limit 
of sensing accuracy set by the random arrival of particles by diffusion at the receptors?
This question was addressed in a seminal work by Berg \& Purcell \cite{berg77}, and
recently reinvestigated by Bialek and Setayeshgar \cite{bialek05,bialek08}. Today, it is generally
accepted that the limit derived by Berg \& Purcell is a fundamental physical limit which cannot be 
exceeded. In this Letter, we show for a single receptor how this limit can be
improved (using maximum likelihood estimation), and discuss how cells could 
implement this improved sensing strategy in practice.\\

Berg \& Purcell calculated the accuracy of concentration sensing by a single receptor
which binds particles of concentration $c_0$ with rate $k_+c_0$ and unbinds particles
with rate $k_-$ (see Fig. 1(a)). Specifically, they considered a binary time series 
of total length $T$ composed of bound and unbound time intervals (see Fig. 1(b)).
Berg \& Purcell estimated concentration directly from the fraction of time $T$ that a particle is bound.
By considering the time correlations of particles bound to the receptor, 
they found the variance $(\delta c)^2$ in the estimated concentration to be \cite{berg77}
\begin{equation}
\frac{(\delta c)^2}{c_0^2}=\frac{2\bar\tau_b}{T\bar p}=\frac{1}{2Dsc_0(1-\bar p)T}\label{eq:BP},
\end{equation}
where $D$ is the diffusion coefficient, $\bar\tau_b$ is the true 
average duration of bound intervals, $s$ describes the receptor dimension, 
and $\bar p$ is the true equilibrium probability for the
receptor to be bound. The last equality in Eq. \ref{eq:BP} is obtained using detailed
balance, {\it i.e.} at equilibrium the rate of unbinding transitions $\bar p/\bar\tau_b$ must equal
the rate of binding transitions $(1-\bar p)/\bar\tau_u$, where $\bar\tau_u$ is the average
duration of unbound intervals. For diffusion-limited binding, $1/\bar\tau_u=4Dsc_0$,
yielding the RHS of Eq. \ref{eq:BP}. In the 
following we revisit the Berg \& Purcell limit on the accuracy of concentration sensing from the 
perspective of maximum likelihood estimation.\\

\begin{figure}[t]
\includegraphics[width=8cm,angle=0]{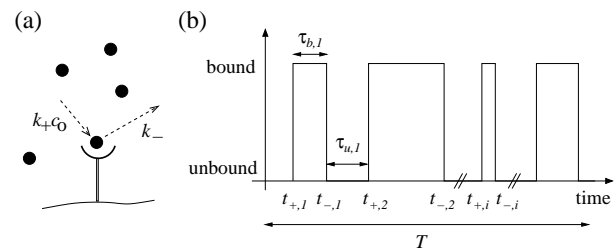}
\caption{\label{fig:fig1} Schematic of particle-receptor binding. (a) An unoccupied receptor can
bind a particle with rate $k_+c_0$, and an occupied receptor can unbind a bound
particle with rate $k_-$. (b) Binary time series of receptor occupancy.}
\end{figure}

Maximum likelihood estimation is a statistical method used for fitting a mathematical model 
to data \cite{ML}. For a fixed set of data and an underlying parameterized model, maximum likelihood picks the 
values of the model parameters that make the data ``more likely'' than they would be for any other values of the 
parameters. Here, the cell's best estimate of concentration can be obtained from maximum likelihood
applied to the time series $\{t_+,t_-\}$ of duration $T$ with particle binding events at times 
$t_{+,i}$ and unbinding events at times $t_{-,i}$ (see Fig. 1(b)). Following Berg \& Purcell,
we disregard potential rebinding of previously bound particles, assuming diffusion is 
sufficiently fast to remove recently unbound particles from the vicinity of the receptor
(but, following \cite{bialek05,bialek08}, we address the more general case in the appendix).

The probability for a time series to occur given a particle concentration $c$ is
\begin{eqnarray}
P(\{t_{+},t_{-}\};c)&=&\nonumber\\
&&\!\!\!\!\!\!\!\!\!\!\!\!\!\!\!\!\!\!\!\!\!\!\!\!\!\!\!\!\!\!\!
\!\!\!\!\!\!\!\!\!\!\!\!\!\!\!\!\!\!\!\!\!\!\!\!\!\!\!\!\!\!\!\!\!
\prod_i p_b(t_{+,i},t_{-,i})p_-(t_{-,i})p_u(t_{-,i},t_{+,i+1})p_+(t_{+,i+1}),\label{eq:P1}
\end{eqnarray}
where the probability for a particle to remain bound from $t_{+,i}$ to $t_{-,i}$ is 
\begin{equation}
p_b(t_{+,i},t_{-,i})=p_b(t_{-,i}-t_{+,i})=e^{-k_-(t_{-,i}-t_{+,i})}
\end{equation}
and the probability for a receptor to remain unbound from $t_{-,i}$ to $t_{+,i+1}$ is
\begin{equation}
p_u(t_{-,i},t_{+,i+1})=p_u(t_{+,i+1}-t_{-,i})=e^{-k_+c(t_{+,i+1}-t_{-,i})}.
\end{equation}
In Eq. \ref{eq:P1}, the probability of binding at time $t_{+,i}$ is 
$p_+(t_{+,i})\propto k_+c$ and the probability of unbinding at time $t_{-,i}$ is 
$p_-(t_{-,i})\propto k_-$. 
Combining all the bound and all the unbound time intervals, we obtain
\begin{equation}
P(\{t_{+,i},t_{-,i}\};c)\propto e^{-k_-T_b}\cdot e^{-k_+cT_u}\cdot k_-^{n}\cdot (k_+c)^{n},\label{eq:P2}
\end{equation}
where $n$ is the number of binding or unbinding events (which can differ by at most 1 
and are therefore approximately equal for $n>\!\!>1$), and $T_{b(u)}=\sum_i^n\tau_{b(u),i}$ 
is the total bound (unbound) time interval with $\tau_{b,i}=t_{-,i}-t_{+,i}$ 
($\tau_{u,i}=t_{+,i+1}-t_{-,i}$).

We maximize $P(\{t_{+},t_{-}\};c)$ over $c$ via
\begin{equation}
\frac{dP}{dc}=-k_+T_uP+\frac{n}{c}P=0,
\end{equation}
and obtain for the maximum likelihood estimate of the particle
concentration
\begin{equation}
\frac{1}{k_+c_\text{\tiny ML}}=\frac{T_u}{n}\quad\text{or}\quad c_\text{\tiny ML}=\frac{n}{k_+T_u}.\label{eq:c_ML}
\end{equation}
Hence, the best estimate of the concentration comes only from the {\it unbound intervals}.
Specifically, $k_+c_\text{\tiny ML}$ is the 
inverse of the average duration of unbound intervals $\tau_u=T_u/n$.
That is,  $k_+c_\text{\tiny ML}$ is just the average binding rate estimated from
the data. 

How accurate is the concentration estimate $c_\text{\tiny ML}$? 
To obtain the uncertainty of the maximum likelihood estimate we require the variance 
$(\delta c_\text{\tiny ML})^2$.
For a given duration $T$ the last interval, possibly an unbound interval, gets interrupted.
To avoid this complication, we consider a fixed number of intervals $n$ (and consequently a 
variable duration $T$) in Eq. \ref{eq:c_ML}. We proceed by using a general relation for the 
variance of the model parameter (here ligand concentration $c$) in maximum likelihood 
estimation. An upper limit of the variance is given by the inverse of the Fisher information 
(Cram\'er-Rao bound)\cite{kay93,shao98}. In our case, 
the Fisher information can be calculated as a simple second derivative of the probability $P$ 
of the data with respect to $c$, averaged over the probability distribution of the
time series at $c_0$. Furthermore, in the limit of a long time series, the Cram\'er-Rao 
bound becomes an equality, and we obtain for the normalized variance
\begin{equation}
\frac{(\delta c_\text{\tiny ML})^2}{c_0^2}=-\frac{1}{c_0^2\left\langle\frac{d^2\ln(P)}{dc^2}\right\rangle_{\!\!c_0}}
=\frac{1}{n},\label{eq:result}
\end{equation}
where we used $P$ from Eq. \ref{eq:P2} \cite{EPAPS}.
Hence, the normalized variance of the maximum likelihood estimate of the true concentration $c_0$ is
exactly the inverse of the number of unbound intervals.

In contrast to our result Eq. \ref{eq:result}, Berg \& Purcell found \cite{berg77} (Eq. \ref{eq:BP})
\begin{equation}
\frac{(\delta c_\text{\tiny BP})^2}{c_0^2}=
\frac{2\bar\tau_b}{T\bar p}=\frac{2\bar\tau_b}{\bar T_b}=
\frac{2}{\bar n_b},\label{eq:BP2}
\end{equation}
where $\bar n_b$ is the average number of bound intervals in the observation time $T$.
Over a long measurement time, the average number of bound and unbound intervals 
must be the same, so the Berg \& Purcell result has exactly twice the variance of the 
maximum likelihood result. 

Why is the maximum likelihood estimate more accurate than the Berg \& Purcell estimate?
Berg \& Purcell assumed that concentration is inferred from the
average bound time, {\it e.g.} as obtained by time averaging the occupancy 
of a single receptor or by spatial averaging over many receptors.
However, as evident from our maximum likelihood estimate, only
the durations of unbound intervals contain information about the concentration. 
In contrast, the average bound time (or equivalently the average unbound time) 
includes the durations of the bound intervals, which add to the uncertainty in estimating the 
concentration.\\

Our result, Eq. \ref{eq:result}, for the variance in the estimate of the concentration $c_0$ 
also predicts optimal binding parameters $k_+$ and $k_-$. Clearly, the more
binding/unbinding events, the lower the variance:\\

\noindent(1) For a given duration $T$ the number of binding/unbinding events is maximized for 
diffusion-limited binding $k_+^\text{max}=4Ds$ (obtained from the diffusive flux $J_\text{max}=4Dsc_0$ 
to an absorbing circular patch of radius $s$).\\

\noindent(2) Similarly, to maximize the number of binding/unbinding events, the unbinding rate $k_-$ should be
maximized. This implies (albeit unrealistically) that $k_-\rightarrow\infty$. \\

Under assumptions (1) and (2), the maximum number of intervals in an observation time $T$ is given by
\begin{equation}
\bar n^\text{max}=\frac{T}{\bar\tau_u^\text{min}}=k_+^\text{max}c_0T=4Dsc_0T,
\end{equation}
leading to a variance
\begin{equation}
\frac{(\delta c_\text{\tiny ML})^2_\text{min}}{c_0^2}=\frac{1}{\bar n^\text{max}}=
\frac{1}{4Dsc_0T}.\label{eq:result3}
\end{equation}
This result can be generalized to the more realistic case of finite $k_-$,
\begin{equation}
\frac{(\delta c_\text{\tiny ML})^2}{c_0^2}=\frac{1}{\bar n}=
\frac{1}{4Dsc_0(1-\bar p)T},\label{eq:result4}
\end{equation} 
where $\bar p=1/(1+k_-/4Dsc_0)$ is the equilibrium probability for the receptor to be bound.
Eq. \ref{eq:result4} can readily be compared with the original Berg \& Purcell result (Eq. \ref{eq:BP}), 
showing again that the maximum likelihood estimate is better by a factor 2.\\

The result for a single receptor, Eq. \ref{eq:result}, can easily be extended to $M$ independent 
receptors, $(\delta c_M)^2/c_0^2=1/(M\bar n)$, {\it i.e.} the variance in the estimated concentration
is the inverse of the total number of unbound intervals for all $M$ receptors. 
However, the concentration estimate cannot become arbitrarily precise with increasing receptor 
number, since the binding of particles is ultimately limited by the arrival of particles 
by diffusion. For an
absorbing circular patch of radius $s'$, particles arrive by diffusion at a rate $4Ds'c_0$. If
individual receptors of effective radius $s$ bind particles at the diffusion-limited rate $4Dsc_0$,
then the number of receptors sufficient to bind all particles incident on the patch is $M\approx s'/s$.
Hence $(\delta c_M)_\text{min}^2/c_0^2\approx 1/[(s'/s)\bar n]$, implying that the variance in the 
concentration estimate decreases at most linearly with the dimension of the detecting surface 
\cite{berg77}.\\


The maximum likelihood concentration estimate Eq. \ref{eq:c_ML} is obtained solely from the duration of 
unbound intervals, thus avoiding the additional uncertainty from the 
bound intervals. What about the alternative scheme of estimating 
the concentration from the {\it number} of binding
events during a time $T$, similar to photon counting by photoreceptors?
As shown below, this estimation scheme approaches the maximum likelihood 
limit as the bound intervals become short.

The average number of binding events (or equivalently bound or unbound intervals) during a time $T$ 
is given by
\begin{equation}
\bar n=\frac{T}{\bar\tau_u+\bar\tau_b}=\frac{T}{\frac{1}{k_+c_0}+\frac{1}{k_-}},
\end{equation}
which provides a concentration estimate $c_\text{est}$ for $c_0$ in terms of the observed $n$,
\begin{equation}
\frac{1}{k_+c_\text{est}}=\frac{T}{n}-\frac{1}{k_-}\quad\text{or}\quad
c_\text{est}=\frac{1}{k_+}\cdot\frac{1}{\frac{T}{n}-\frac{1}{k_-}}.\label{eq:c_est}
\end{equation}
From the standard deviation of $n$, we obtain the standard deviation of $c_\text{est}$ via
\begin{equation}
\delta c_\text{est}=\frac{dc_\text{est}}{dn}\delta n\label{eq:c_est2}.
\end{equation}
According to Eq. \ref{eq:c_est}, the derivative $dc_\text{est}/dn$ is given by
\begin{equation}
\frac{dc_\text{est}}{dn}=\frac{k_+T}{n^2}c_\text{est}^2\approx\frac{k_+T}{\bar n^2}c_0^2.\label{eq:dcdn}
\end{equation}
To obtain $\delta n$ for a fixed duration $T$, we note that this
is proportional to the standard deviation $\delta T$ for fixed $n$ via 
$\delta n=(dn/dT)\delta T$. 
Using $\bar T=n(\bar\tau_u+\bar\tau_b)$, yields $dT/dn=\bar\tau_u+\bar\tau_b$, leading to
${dn/dT}=1/(\bar\tau_u+\bar\tau_b)$.
Based on the variance of unbound (bound) intervals 
$\langle(\tau_{u(b)}-\bar\tau_{u(b)})^2\rangle=\bar\tau_{u(b)}^2$, calculated
from $\bar \tau_{u(b)}=\langle\tau_{u(b)}\rangle$ and 
$\langle\tau_{u(b)}^2\rangle=(1/\bar\tau_{u(b)})\int_0^\infty dt\, t^2 \exp{(-t/\bar\tau_{u(b)})}=2\bar\tau_{u(b)}^2$, 
we obtain for
\begin{equation}
(\delta T)^2=n(\bar\tau_u^2+\bar\tau_b^2)\quad\text{or}\quad\delta T
=\sqrt{n(\bar\tau_u^2+\bar\tau_b^2)}.\label{eq:dT}
\end{equation}
Finally, using these results in Eq. \ref{eq:c_est2} leads to
\begin{equation}
\frac{(\delta c_\text{est})^2}{c_0^2}=\frac{1+\left(\frac{\bar\tau_b}{\bar\tau_u}\right)^2}{n}.
\end{equation}
This variance interpolates between the maximum likelihood and the 
Berg \& Purcell results for $\bar\tau_b<\bar\tau_u$
and exceeds the Berg \& Purcell limit for $\bar\tau_b>\bar\tau_u$. To provide some intuition for this result,
we consider two limits:\\

\noindent(1)  $\bar\tau_b<\!\!<\bar\tau_u$: In this regime, the brief bound intervals do not contribute 
appreciably to $T$. As a result, counting the number of binding events in a time $T$ is the 
same as estimating the mean unbound time interval $\bar \tau_u$. This is exactly the maximum likelihood 
estimator (Eq. \ref{eq:c_ML}).\\

\noindent(2)  $\bar\tau_b>\!\!>\bar\tau_u$: In this regime, the bulk of the time $T$ is accounted for by 
the bound intervals. Therefore, the number of binding events measures the duration of the bound intervals, 
not the duration of the unbound intervals, which contain all the
information about the concentration. (The Berg \& Purcell estimate is more accurate in this regime 
because the fraction of time spent bound effectively measures the ratio of the bound to unbound time, and therefore
captures information about the duration of unbound intervals.)\\

Our analysis has neglected additional noise in the concentration estimate due to ligand rebinding 
(\cite{bialek05}, also see Appendix A). However, cells have mechanisms for eliminating ligands which 
could suppress this noise \cite{endres08,endres09}. Examples include ligand-receptor 
internalization \cite{ferguson01,schandel94}, and enzymatic degradation of ligands, {\it e.g.} of cAMP 
ligand by membrane bound phosphodiesterases in {\it Dictyostelium discoideum}  \cite{sucgang97}. 
In fact, internalization can be very efficient; the transferrin receptor (TfR)
and the low-density lipoprotein receptor (LDLR) are internalized, respectively, 6.7 and 4.9 times
faster than their specific ligands can unbind \cite{shankaran07}.

With or without ligand rebinding, to what extent can real cells exploit any 
of the above maximum likelihood schemes to improve the accuracy of concentration sensing? It is not
clear mechanistically how cells could sense and respond exclusively to the durations of unbound
intervals (Eq. \ref{eq:c_ML}). The potentially more practical scheme in Eq. \ref{eq:c_est} of counting the number
of binding events in a time $T$ can approach the maximum likelihood limit for $\bar\tau_b<\!\!<\bar\tau_u$
(though too short a bound interval $\tau_b$ might imply low ligand specificity \cite{feinerman08} 
and potential signaling crosstalk). Effective counting can be achieved by receptor adaptation or desensitization 
following ligand binding. An intriguing alternative is that receptors could bind ligand once and then
be internalized before ligand is released.
While it is an open question whether cells actually implement this ``optimal'' strategy, we hope the perspective 
provided by maximum likelihood will prove useful in interpreting some of the complexities of cellular signaling
systems.\\

We thank Pankaj Mehta for valuable suggestions. R.G.E. acknowledges funding from the
Biotechnology and Biological Sciences Research Council grant BB/G000131/1 
and the Centre for Integrated Systems Biology at Imperial College (CISBIC). 
N.S.W. acknowledges funding from the Human Frontier Science Program (HFSP) 
and the National Science Foundation grant PHY-0650617.\\

\appendix

\section{\label{secA} Maximum likelihood estimate with rebinding }

Following Berg \& Purcell, our derivation neglected the rebinding of already measured particles.
Such rebinding increases the uncertainty in estimating the concentration \cite{bialek05,endres08}.
As rebinding noise can be avoided by ligand-receptor 
internalization or ligand degradation on cell surfaces \cite{endres08,endres09}, 
it does not contribute to the fundamental physical limit. However, in practice many
receptors do release and potentially rebind their ligands.

The effect of local particle diffusion and hence possible rebinding is to make the instantaneous rate 
of binding a functional of the previous binding and unbinding events (see Ref. \cite{bialek05} 
for details). The binding rate can thus be written as $k_+c(t,\{t_{+},t_{-}\})$. The rate of 
unbinding remains $k_-$, so the maximum likelihood estimate of concentration still comes entirely 
from the durations of the unbound intervals. 

What is the maximum likelihood estimate $c_\text{\tiny ML}$?
The probability for a time series is still given by Eq. \ref{eq:P1} with
the change due to diffusion and rebinding occurring in $p_+\propto k_+(c+\Delta c_i)$ and $p_u$:
\begin{equation}
p_u(t_{-,i},t_{+,i+1})=e^{-k_+c(t_{+,i+1}-t_{-,i})-k_+\int_i \Delta c(t')dt'},\label{eq:pu}
\end{equation}
where we have expressed the particle concentration as
\begin{eqnarray}
c(t,\{t_+,t_-\})&=&c+\Delta c(t,\{t_{+},t_{-}\})\nonumber\\
&=&c+\Delta c(\{t-t_{-,i};t-t_{+,i}\}),
\end{eqnarray}
and used the notation $\int_idt'=\int_{t_{-,i}}^{t_{+,i+1}}dt'$,
$\Delta c(t')=\Delta c(t',\{t_+,t_-\})$, and $\Delta c_i=\Delta c(t_{+,i})$.

The terms can be gathered as before, leading to
\begin{eqnarray}
P(\{t_{+},t_{-}\};c)&\propto&e^{-k_-T_b}\cdot e^{-k_+cT_u}\cdot k_-^{n}\cdot k_+^{n}\nonumber\\
&&\!\!\!\!\!\!\!\!\!\!\!\!\!\!\!\!\!\!\!\!
\cdot\prod_i(c+\Delta c_i)e^{-k_+\int_i\Delta c(t') dt'}.
\end{eqnarray}
Importantly, all the $\Delta c$'s depend only on the times of events, not the value of $c$, so 
$d(\Delta c)/dc=0$, yielding
\begin{equation}
\frac{dP}{dc}\propto-k_+T_uP+\sum_i\frac{1}{c+\Delta c_{i}}P.
\end{equation}
Setting the above derivative to zero yields an implicit equation for the maximum likelihood estimate
$c_\text{\tiny ML}$,
\begin{equation}
\sum_i\frac{1}{c_\text{\tiny ML}+\Delta c_i}=k_+T_u,
\end{equation}
where the sum is over all binding events, but each $\Delta c_i$ depends deterministically 
on all previous binding and unbinding events. Using again that the variance of a maximum likelihood 
estimator is given by the inverse of the Fisher information \cite{kay93,shao98}, we obtain
\begin{equation}
\frac{(\delta c_\text{\tiny ML})^2}{c_0^2}=\frac{1}{\sum_i(1+\Delta c_i/c_0)^{-2}}.
\end{equation}

\section{\label{secB} Alternative derivation of maximum likelihood estimate without particle rebinding}

Here we obtain the uncertainty of the maximum likelihood estimate $c_\text{\tiny ML}$ 
of the single receptor without particle rebinding by directly
calculating the variance $(\delta c_\text{\tiny ML})^2$ explicitly (not using the 
inverse of the Fisher information). As in the main text,
we consider a fixed number of intervals $n$ (and consequently a 
variable duration $T$).

The ensemble average of Eq. 7 in the main text is simply given by
\begin{equation}
\left\langle\frac{1}{k_+c_\text{\tiny ML}}\right\rangle=\frac{\langle T_u\rangle}{n}
=\bar\tau_u=\frac{1}{k_+c_0}.
\end{equation}
To obtain the variance $(\delta c_\text{ML})^2$, we proceed by calculating
\begin{widetext}
\begin{equation}
\left\langle\left(\frac{1}{k_+c_\text{\tiny ML}}\right)^2\right\rangle=\langle\tau_u^2\rangle=
\bar\tau_u^2\left\langle\left( 1+\frac{\frac{1}{n}\sum_i^n(\tau_{u,i}-\bar\tau_u)}{\bar\tau_u}\right)^2\right\rangle=
\bar\tau_u^2\left(1+\frac{\sum_i^n\langle(\tau_{u,i}-\bar\tau_u)^2\rangle}{n^2\bar\tau_u^2}\right)=
\bar\tau_u^2\left(1+\frac{\langle(\tau_{u}-\bar\tau_u)^2\rangle}{n\bar\tau_u^2}\right),\label{eq:wide}
\end{equation}
\end{widetext}
where we used the definition $\bar \tau_u=\langle\tau_u\rangle$ and the fact that
the durations of the different unbound intervals are uncorrelated.
Using $\langle\tau_u^2\rangle=(1/\bar\tau_u)\int_0^\infty dt\, t^2 \exp{(-t/\bar\tau_u)}=2\bar\tau_u^2$, 
we obtain for the variance of unbound intervals $\langle(\tau_{u}-\bar\tau_u)^2\rangle=\bar\tau_u^2$,
which substituted into Eq. \ref{eq:wide} yields
\begin{equation}
\left\langle\left(\frac{1}{k_+c_\text{\tiny ML}}\right)^2\right\rangle=
\bar\tau_u^2\left(1+\frac{1}{n}\right).\label{eq:E_squared}
\end{equation}
By subtracting $\langle 1/(k_+c_\text{ML})\rangle^2=\bar\tau_u^2$ from Eq. \ref{eq:E_squared}, we obtain the variance 
\begin{equation}
\left\langle\left(\frac{1}{k_+c_\text{\tiny ML}}\right)^2\right\rangle-
\left\langle\frac{1}{k_+c_\text{\tiny ML}}\right\rangle^2=
\frac{\bar\tau_u^2}{n}.\label{eq:var}
\end{equation}
Then using
\begin{equation}
\delta\left(\frac{1}{k_+c_\text{\tiny ML}}\right)=-\frac{\delta c_\text{\tiny ML}}{k_+c_\text{\tiny ML}^2}
\approx -\frac{\delta c_\text{\tiny ML}}{k_+c_0^2},
\end{equation}
which is valid for small relative standard deviation $\delta c_\text{\tiny ML}/c_0$ in Eq. \ref{eq:var},
we obtain the same result as Eq. 8 of the main text, {\it i.e.}
\begin{equation}
\frac{(\delta c_\text{\tiny ML})^2}{c_0^2}=
k_+^2c_0^2\left\langle\delta\left(\frac{1}{k_+c_\text{\tiny ML}}\right)^2\right\rangle=
\frac{(k_+c_0\bar\tau_u)^2}{n}=\frac{1}{n}.\label{eq:result}
\end{equation}


\noindent $*\ $r.endres@imperial.ac.uk\\
\noindent $\dagger\ $wingreen@princeton.edu\\

\end{document}